\pdfoutput=1
\documentclass[]{spie}  

\usepackage{amsmath,amsfonts,amssymb}
\usepackage{graphicx}
\usepackage[colorlinks=true, allcolors=blue]{hyperref}

\title{Development of high-resolution arrayed waveguide grating spectrometers for astronomical applications: first results}

\author[a]{Pradip Gatkine} 
\author[a,b]{Sylvain Veilleux}
\author[c]{Yiwen Hu}
\author[c]{Tiecheng Zhu}
\author[c]{Yang Meng}
\author[d]{Joss Bland-Hawthorn}
\author[c]{Mario Dagenais}

\affil[a]{Department of Astronomy, University of Maryland, College Park, Maryland 20742, USA}
\affil[b]{Joint Space-Science Institute, University of Maryland, College Park, Maryland 20742, USA}
\affil[c]{Department of Electrical and Computer Engineering, University of Maryland, College Park, Maryland 20742, USA }
\affil[d]{Sydney Institute for Astronomy and Sydney Astrophotonic Instrumentation Labs, School of Physics, The University of Sydney, New South Wales 2006, Australia}

\authorinfo{Further author information: (Send correspondence to P. Gatkine)\\E-mail: pgatkine@astro.umd.edu}

\begin{document} 
\maketitle

\begin{abstract}
Astrophotonics is the next-generation approach that provides the means to miniaturize near-infrared (NIR) spectrometers for upcoming large telescopes and make them more robust and inexpensive. The target requirements for our spectrograph are: a resolving power of $\sim$3000, wide spectral range (J and H bands), free spectral range of about 30 nm, high on-chip throughput of about 80\% (-1dB) and low crosstalk (high contrast ratio) between adjacent on-chip wavelength channels of less than 1\% (-20 dB). A promising photonic technology to achieve these requirements is Arrayed Waveguide Gratings (AWGs). We have developed our first generation of AWG devices using a silica-on-silicon substrate with a very thin layer of Si$_3$N$_4$ in the core of our waveguides. The waveguide bending losses are minimized by optimizing the geometry of the waveguides. Our first generation of AWG devices are designed for H band have a resolving power of $\sim$1500 and free spectral range of $\sim$ 10 nm around a central wavelength of 1600 nm. The devices have a footprint of only 12 mm $\times$ 6 mm. They are broadband (1450-1650 nm), have a peak on-chip throughput of about 80\% ($\sim$ -1 dB) and contrast ratio of about 1.5\% (-18 dB). These results confirm the robustness of our design, fabrication and simulation methods. Currently, the devices are designed for Transverse Electric (TE) polarization and all the results are for TE mode. We are developing separate J- and H-band AWGs with higher resolving power, higher throughput and lower crosstalk over a wider free spectral range to make them better suited for astronomical applications.

\end{abstract}


\section{INTRODUCTION}
\label{sec:intro}  

The J- and H-bands of near-infrared (NIR) light are crucial in studying phenomena in the first billion years (z $\sim$ 6-12) of the universe such as galaxy formation, ionization of intergalactic medium and supermassive black hole formation. In this range of redshifts, the rest-frame UV light (specifically Lyman-$\alpha$), characteristic of the star formation, is redshifted to NIR band. To obtain the NIR spectra of high redshift probes such as quasars and gamma ray burst (GRB) afterglows, large telescopes such as Keck 10-meter telescopes are required.

\noindent
The next-generation of ground-based extremely large telescopes (ELTs) in optical and NIR will have diameters in the range of thirty meters (eg. Thirty Meter Telescope). This necessitates the development of suitable seeing limited spectroscopic instrumentation \cite{bland2006instruments}. The size of the optical components in a  conventional spectrograph scales roughly with the telescope diameter D, hence the volume, mass, and cost of the instrument scale roughly as diameter cubed \cite{bland2006instruments}. This highlights the need for innovation to build instruments for upcoming Extremely Large Telescopes (ELTs). Astrophotonics is a new approach that will miniaturize the next-generation spectrometers for large telescopes by the virtue of its two-dimensional structure \cite{bland2009astrophotonics}. As each pixel is a fiber at the slit, the collimating lenses are no longer required. The light is guided through the fibers and waveguides into the 2-dimensional structure for dispersion, thus reducing the size of spectroscopic instrumentation to few centimeters and the weight to a few hundreds of grams. These devices are also much less expensive than conventional astronomical spectrographs with commensurate specifications (resolution, efficiency and operating wavelength range) \cite{cvetojevic2010miniature}.    

\subsection{Arrayed Waveguide Gratings (AWGs)}
AWG device has originated from the need of increasing the data rate in the field of fiber-optic communication. This was achieved by using AWG as a ‘wavelength separating’ and ‘wavelength combining’ device. The pioneering paper by Meint Smit \cite{smit1996phasar} described the detailed theory of AWG design in 1996. The traditional uses of AWG involved high power sources centered on a narrow band (5-10 nm) around wavelength 1550 nm, which is the standard in telecommunication industry. But in principle, the same theory can be used for spectroscopic purposes. In particular, some of the recent work towards making ultra-low loss AWG devices \cite{bauters2010ultra, dai2011low, akca2011high} made it possible to explore the utility of AWG based spectrographs in astronomical instrumentation. There have also been successful preliminary tests of using modified commercial AWGs for astronomical spectroscopy \cite{cvetojevic2012developing, cvetojevic2012first}. AWGs, along with other advances in the field of astrophotonics, such as photonic lanterns \cite{leon2010photonic} to convert multimode fibers to single mode fibers,  Bragg gratings (in-fiber \cite{trinh2013gnosis} as well as on-chip \cite{zhu2016arbitrary}) to suppress unwanted OH-emission background (in NIR) and high-efficiency fiber bundles for directly carrying the light from the telescope focal plane \cite{lawrence2012hector}, offer a complete high efficiency miniaturized solution for astronomical spectroscopy in NIR. This has a potential to be a paradigm shifting development for future ground-, balloon- and space-based telescopes. 

\begin{figure} [ht]
   \begin{center}
   \begin{tabular}{c} 
   \includegraphics[height=6cm]{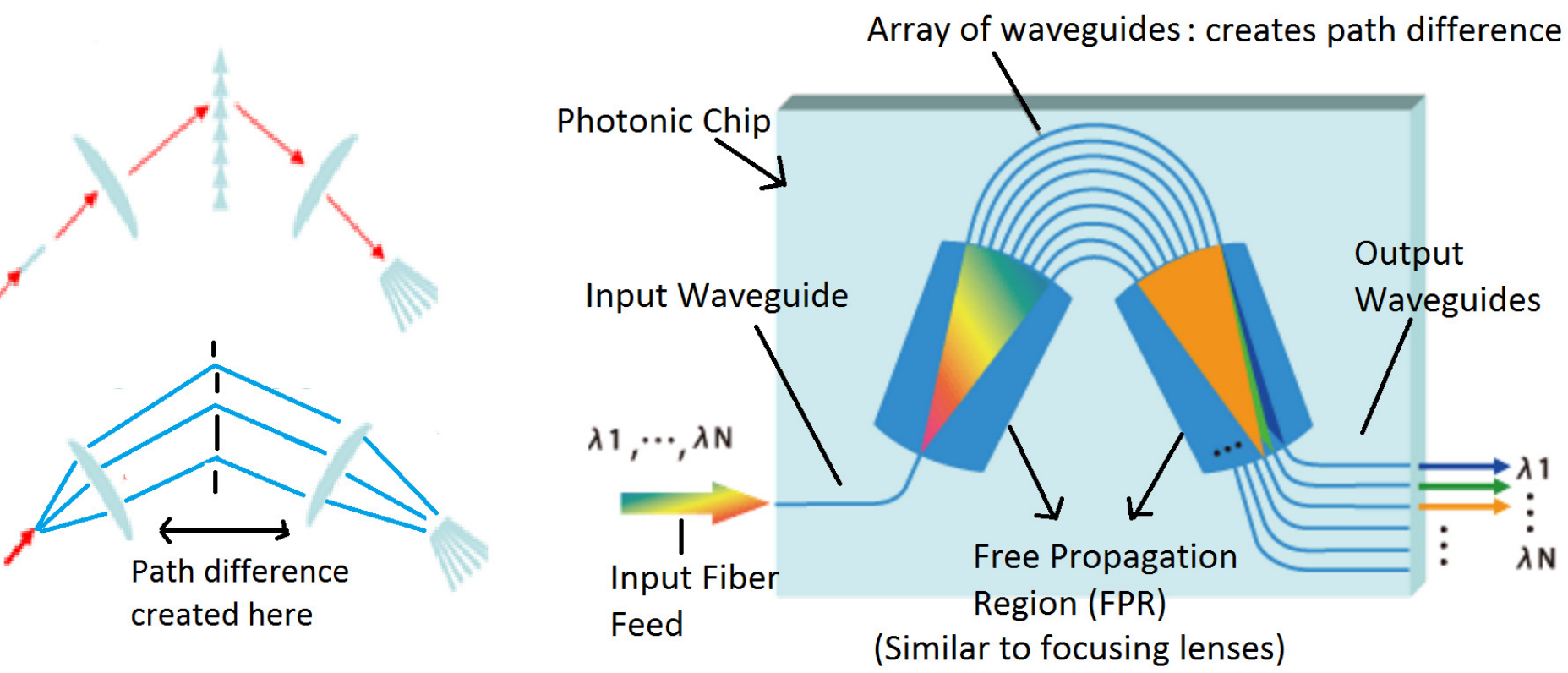}
   \end{tabular}
   \end{center}
   \caption[AWG_sim_conventional] 
   { \label{fig:AWG_sim_conventional} 
Analogy between conventional grating spectrograph (left) and arrayed waveguide gratings (right)}
   \end{figure}

\begin{figure} [ht]
   \begin{center}
   \begin{tabular}{c} 
   \includegraphics[height=4.5cm]{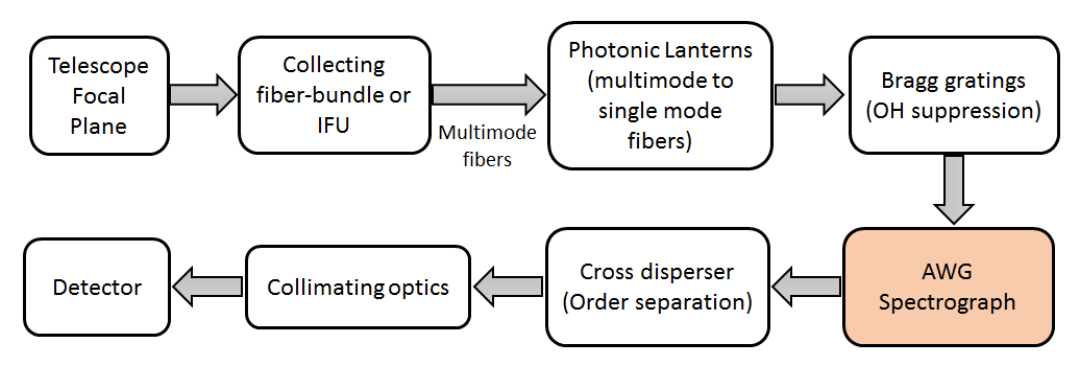}
   \end{tabular}
   \end{center}
   \caption[AWG_telescope] 
   { \label{fig:AWG_telescope} 
A sample light path with AWG used as a spectrograph in an astronomical setup}
   \end{figure}

\noindent
The analogy between AWGs and a conventional grating spectrograph is shown in Fig.~\ref{fig:AWG_sim_conventional}. Light propagates in confined guided paths by the principle of total internal reflection. The focusing lenses are replaced by on-chip free propagation regions (FPR) and the gratings are replaced by the array of waveguides guiding the light along the lines with a fixed desired path differences by linearly increasing the waveguide length from bottom to top. The fixed path difference between adjacent waveguides depends on the spectral order. The output light from the array propagates through the output FPR and different wavelengths constructively interfere at different positions on the concave output facet. The dispersed light can be collected in output waveguides or sent to the detector through free space. A sample light path with AWG used as a spectrograph is shown in Fig.~\ref{fig:AWG_telescope}.

\subsection{Target Specifications}
The requirements for the spectrograph are set by the science goals. The main science goals we are considering here are: (1) precise measurement of the metallicity of high-z GRB hosts and the intervening systems by measuring the equivalent widths of a few metal-diagnostic lines and (2) a proper characterization of red damped wing of Lyman-$\alpha$ in GRB afterglow spectrum for constraining HI column density (N$_H$$_I$) and intergalactic medium properties. Achieving these goals requires a spectral resolving power ($\lambda$/$\Delta\lambda$) of $\sim$3000 as well as a wide spectral range (J and H band) so that these lines fall within the band-pass for GRBs at redshifts of $\sim$6 to 12. The throughput of the spectrograph should be commensurate with typical conventional spectrograph throughput (from slit to detector) of about 20\%, which translates to an on-chip efficiency of about 80\% (due to additional coupling losses). The total efficiency of an AWG spectrograph is the product of two parameters: 
(1) On-chip efficiency (efficiency from start of input waveguide to the end of output waveguides on the chip) and 
(2) Coupling efficiency (efficiency of coupling from input fiber to input waveguide and output waveguide to output fiber).   

For the preliminary development of our astronomical AWG spectrograph, we relax the stringent specifications to first validate and establish a robust procedure for design, fabrication and characterization. The preliminary and future target specifications for the AWG are stated in Table 1. For the rest of this paper we will be focusing on preliminary target specifications.   

\begin{table}[ht]
\caption{Preliminary and future target specifications for the proposed AWG device in the order of importance} 
\label{tab:Target_specs}
\begin{center}       
\begin{tabular}{|l|c|c|} 
\hline
\rule[-1ex]{0pt}{3.5ex} \textbf{Parameters} & \textbf{Preliminary Target} & \textbf{Future Target} \\
\hline
\rule[-1ex]{0pt}{3.5ex}  1. On-chip Throughput & 80\% & 90\%   \\
\hline
\rule[-1ex]{0pt}{3.5ex} \begin{tabular}[t]{@{}l@{}}2. Peak Overall Throughput\\(including coupling + propagation)\end{tabular} & 15\% & 35\%   \\
\hline
\rule[-1ex]{0pt}{3.5ex}  3. Operating Waveband & H band (1450-1700 nm)  &   \begin{tabular}[t]{@{}c@{}}H band (1450-1700 nm)\\J-band (1150-1400 nm)\end{tabular} \\
\hline
\rule[-1ex]{0pt}{3.5ex}  4. Resolving Power (R) & $\sim$1500 & $\sim$3000  \\
\hline
\rule[-1ex]{0pt}{3.5ex} \begin{tabular}[t]{@{}c@{}}5. Adjacent Crosstalk\\($\sim$ contrast ratio)\end{tabular} & $\sim$1\% & $\sim$1\%  \\
\hline 
\rule[-1ex]{0pt}{3.5ex}  6. Free Spectral Range & 10 nm & 30 nm  \\
\hline

\end{tabular}
\end{center}
\end{table}
\noindent

\section{Methods}
\label{sec:methods}  
The target specifications in astronomy are very different from the traditional narrowband, high power applications of AWGs in telecommunication industry from where the concept of AWGs originate. The book by Okamoto \cite{okamoto2010fundamentals} explains the principles of operation and fundamental characteristics of AWGs. The pioneering paper by Meint Smit \cite{smit1996phasar} dwells upon the design procedures, geometrical layout and various practical issues of AWGs. The first theoretical and experimental work towards proving the capabilities and limitations of an astronomical AWG spectrograph was conducted by Lawrence et al. \cite{lawrence2010miniature}. Buried silicon nitride platform is the leading solution for low-loss AWGs. The photonics group at UC Santa Barbara have performed several experiments with this platform for applications in telecommunication industry \cite{bauters2010ultra, dai2011low}. These results are also useful for the development of an astronomical AWG spectrometer. 

\noindent
Developing a photonic AWG spectrometer involves 4 broad steps: design, simulation, fabrication and characterization of the instrument. In our first design we are developing an AWG spectrometer for H band (1450-1650 nm). In this section, we will describe the design methodology followed, the simulations performed and fabrication techniques developed to achieve our preliminary target specifications.

\subsection{Design}
\label{Design}
\begin{figure} [ht]
   \begin{center}
   \begin{tabular}{c} 
   \includegraphics[height=4cm]{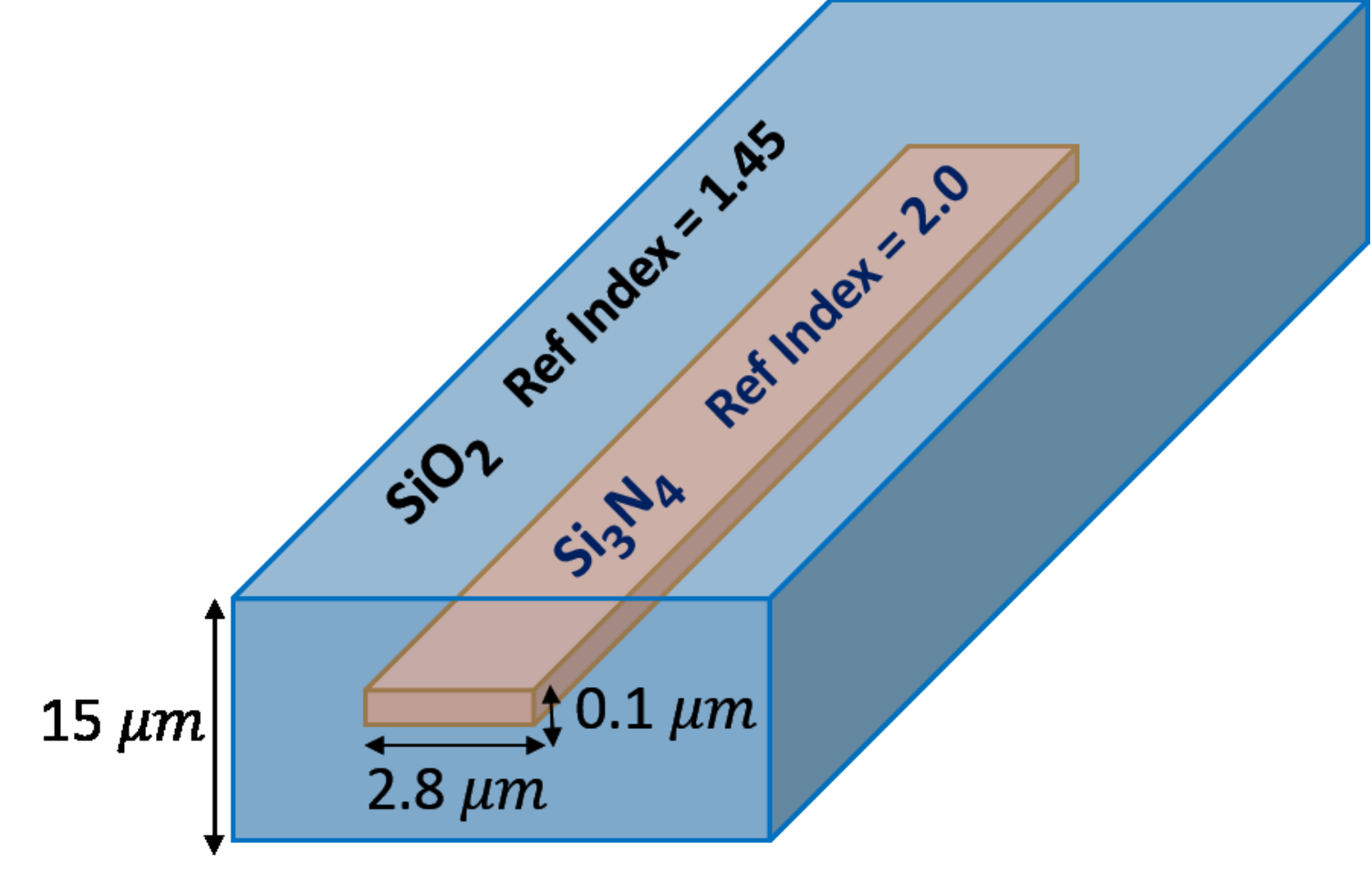}
   \end{tabular}
   \end{center}
   \caption[Waveguide_geometry] 
   { \label{fig:Waveguide_geometry} 
Geometry and material profile of the waveguide cross-section. }
   \end{figure}

For developing low-loss AWGs, the selection of material profile is crucial. We use Si$_3$N$_4$ waveguides buried in SiO$_2$ as shown in Fig.~\ref{fig:Waveguide_geometry} since it is the most suitable configuration for low on-chip transmission loss\cite{bauters2010ultra}. The most important sources of loss in AWG (in the order of importance) are: 
\\
\\* 1. \textbf{Coupling loss} from fiber to waveguide and vice versa due to mismatch between fiber mode and waveguide mode 
\\* 2. \textbf{Sidewall scattering loss}: The fabrication processes limit the smoothness of sidewalls of the waveguides, thus leading to a loss of guided light through scattering from the sidewalls. This loss can be mitigated by making very thin waveguides to reduce the sidewall area. 
\\* 3. \textbf{Bending loss} due to radiation-leaking of light from the curved waveguides in AWG. A weakly confined mode loses light more easily than a strongly confined mode. 
\\* 4. \textbf{Absorption loss}: The waveguide material has a finite absorption co-efficient which leads to absorption of a small amount of guided light.

\noindent
In our first design we focused on sidewall loss and bending loss. We used a waveguide width of 2.8$\mu$m and a height of 0.1$\mu$m in order to minimize the bend loss as demonstrated in Ref\cite{bauters2010ultra} and sidewall roughness loss\cite{dai2011low}, since AWGs have extended curved waveguides. This waveguide geometry is shown in Fig.~\ref{fig:Waveguide_geometry}. The waveguide geometry and the design was modified in the second version for better coupling efficiency (as explained in results section). 

\begin{figure} [ht]
   \begin{center}
   \begin{tabular}{c} 
   \includegraphics[height=5cm]{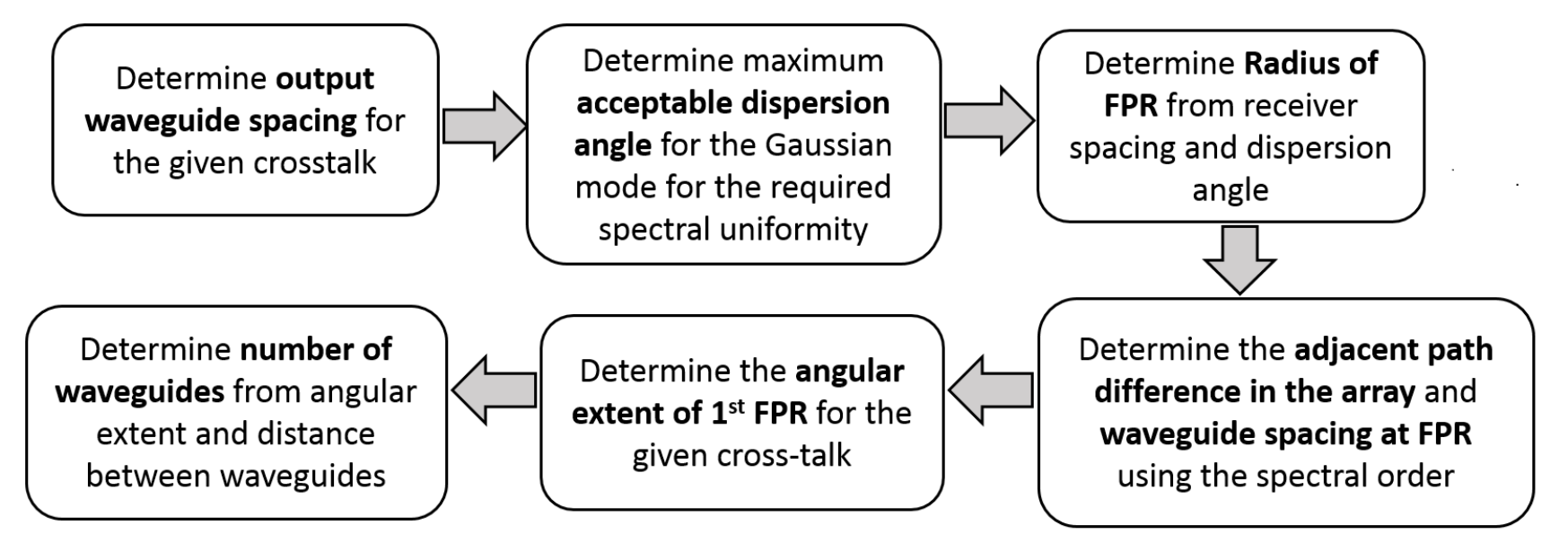}
   \end{tabular}
   \end{center}
   \caption[Design_flow] 
   { \label{fig:Design_flow} 
AWG Design procedure. For details, please see Ref\cite{smit1996phasar}}
   \end{figure}

\noindent
The dispersed light for all the spectral orders is received by the ‘output channels’ (i.e. output waveguides) and guided to the edge of the chip, called $‘$facet$’$, where it can be coupled with the output fiber for characterization. We chose to sample the focal plane of dispersed light at 5 discrete uniformly spaced points to ensure sufficient characterization. Thus the design has 5 output waveguides, each sampling a specific region of the output plane and the output spectrum. We developed a code to automate the process of design as shown in Fig.~\ref{fig:Design_flow}.

\noindent
\textbf{Selection of spectral order}: 
\vspace{1mm}
\\* A spectral order m = 165 (at $\lambda$ = 1600 nm) is used in this design to maximize the free spectral range and at the same time keeping the incremental path difference between adjacent arrayed waveguide and thereby, the size of the chip small. Using a higher spectral order results into a higher incremental path difference and increased chip-size. Also, if the order is too small, the center to center distance between adjacent waveguides at the free propagation region (FPR) interface reduces, thus making the FPR overcrowded and leading to higher crosstalk between adjacent waveguides. Therefore, a spectral order m = 165 is used to strike a balance between free spectral range (FSR) and chip-size for the first design. In future, we plan to push the boundaries to increase FSR to the target specification of 30 nm.

\subsection{Simulations}
\begin{figure} [ht]
   \begin{center}
   \begin{tabular}{c} 
   \includegraphics[height=5cm]{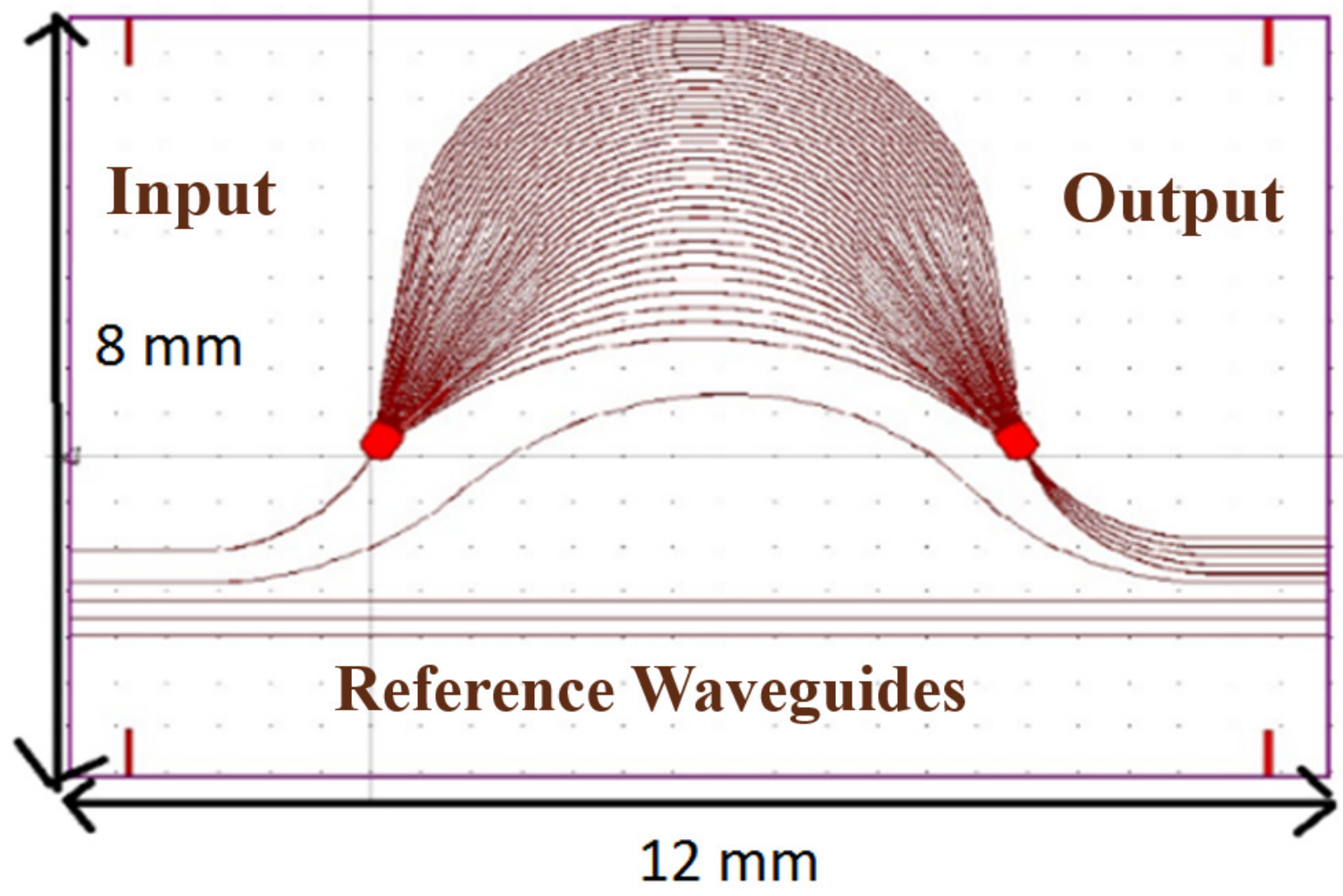}
   \end{tabular}
   \end{center}
   \caption[AWG_CAD] 
   { \label{fig:AWG_CAD} 
CAD of AWG (first design). Note the vertical cleaving marks near the four corners to aid cleaving the edges to expose optical quality cross-section of waveguides for fiber coupling. The extra waveguides at the bottom are for calibration}
   \end{figure}

\begin{figure} [ht]
   \begin{center}
   \begin{tabular}{c} 
   \includegraphics[height=5cm]{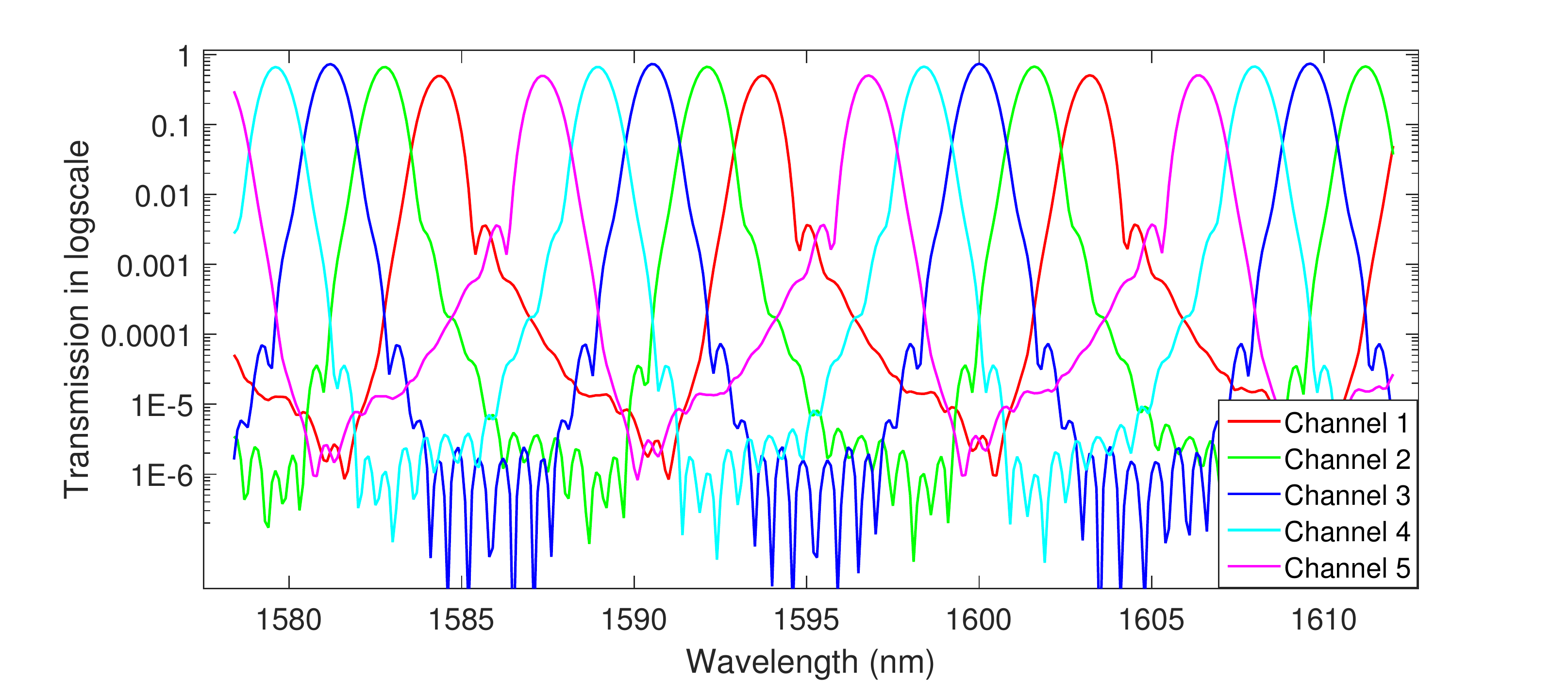}
   \end{tabular}
   \end{center}
   \caption[Simulation_results-eps-converted-to.pdf] 
   { \label{fig:Simulation_results} 
 Simulated transmission of 5 output channels of AWG (4 spectral orders shown here) }
   \end{figure}

We simulate the design obtained in the step above using RSoft CAD\cite{rsoft200layout} and AWG utility of BeamPROP software\cite{BeamPROP}. The CAD for the first AWG design is shown Fig.~\ref{fig:AWG_CAD} and the simulation result is shown in Fig.~\ref{fig:Simulation_results}. There are straight and curved reference waveguides below the AWG (in the CAD), which are used to calibrate the output. The overall size of the AWG chip is 8mm $\times$ 12mm. There is one input waveguide and 5 output waveguides for this AWG. In the spectral domain, these output waveguides are separated by  $\Delta$$\lambda$ $\sim$ 1.6nm in all the spectral orders. This is also called ‘channel spacing’. Note that the results of the simulations are very close to the target specifications. 

\subsection{Fabrication}
\begin{figure} [ht]
   \begin{center}
   \begin{tabular}{c} 
   \includegraphics[height=3.5cm]{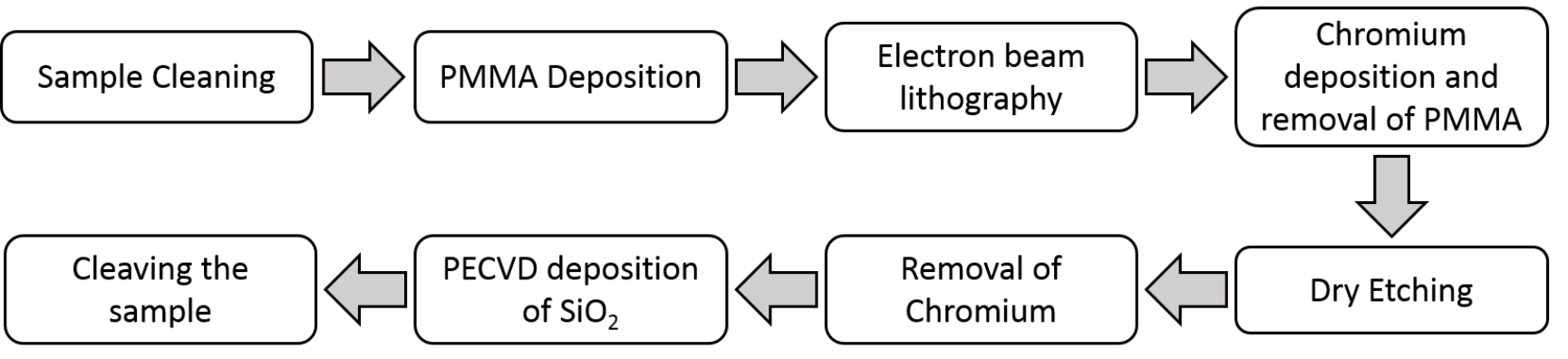}
   \end{tabular}
   \end{center}
   \caption[Fabrication_process] 
   { \label{fig:Fabrication_process} 
 Fabrication sequence for AWG. Here, PMMA is a chemical used as photoresist (a material which changes property when it is exposed to electron beam) and is washed out when the exposed area is developed with a chemical developer (called MIBK).}
   \end{figure}

To fabricate AWG chip, we use silicon substrate with 10$\mu$m of thermal silica (SiO$_2$) pre-deposited on it. We deposit 0.1$\mu$m silicon nitride (Si$_3$N$_4$) on top of that using LPCVD (Low Pressure Plasma Enhanced Deposition). The processes followed to fabricate buried Silicon Nitride AWG device (in the sequence) are: Electron-beam lithography, electron-beam Chromium metal deposition, Chromium lift-off leaving only the Chromium mask for etching, reactive ion etching (RIE) to a depth of 0.1$\mu$m, Chromium etching to dissolve the mask and finally, PECVD (Plasma Enhanced Chemical Vapor Deposition) of 6$\mu$m of SiO$_2$ as the upper cladding layer of the device. After fabrication, the sample is cleaved at precise locations from left end and right end to expose facets of the input and output waveguides for coupling the light. The facets are of optical quality, since the sample breaks along the crystal plane of the chip. The fabrication sequence is briefly shown in Fig.~\ref{fig:Fabrication_process}.     

\subsection{Characterization}
\begin{figure} [ht]
   \begin{center}
   \begin{tabular}{c} 
   \includegraphics[height=5.5cm]{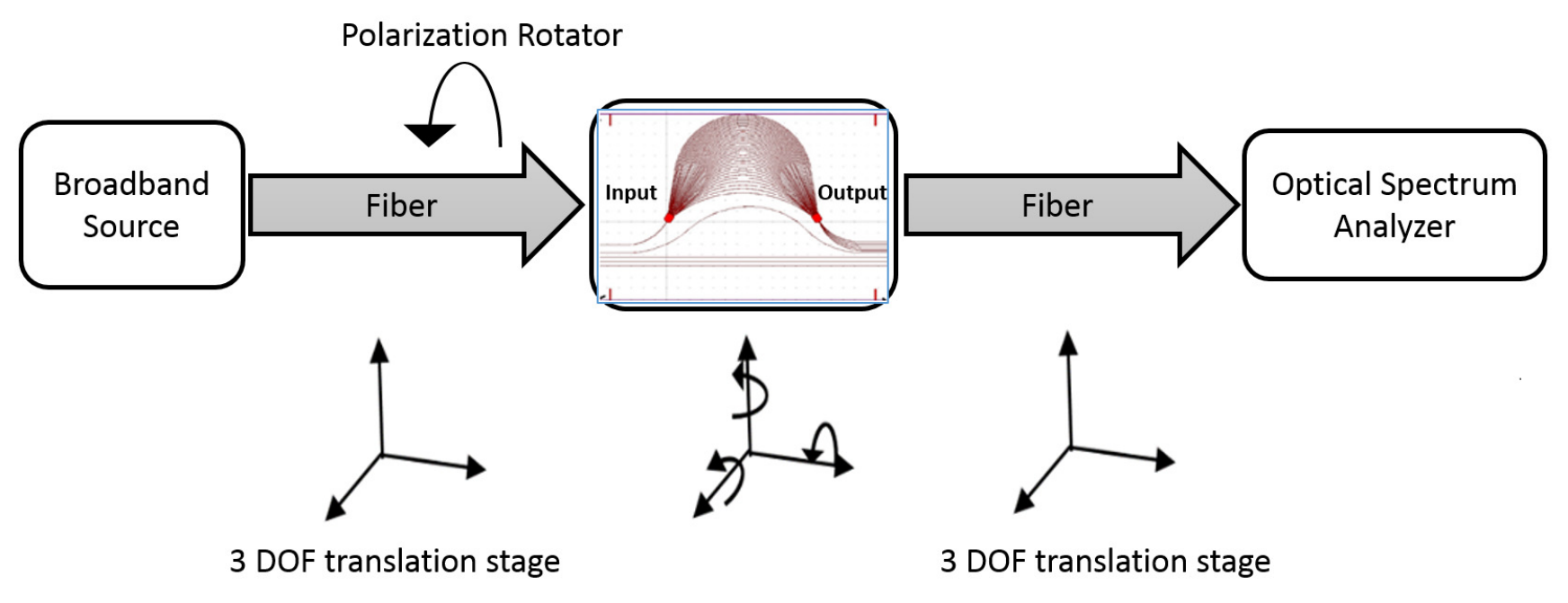}
   \end{tabular}
   \end{center}
   \caption[Characterization_stage] 
   { \label{fig:Characterization_stage} 
 A schematic of the characterization setup with its main constituents and degrees of freedom.}
   \end{figure}

To characterize the transmission spectrum of the AWG, a broadband superluminescent diode source by Throlabs (S5FC1550P- A2) operating in a waveband of 1450 nm - 1650 nm is used. A polarization maintaining (PM) fiber, PM1550-XP is connected to this light source and is optically coupled to the AWG input waveguide. Another PM fiber is optically coupled to one of the output waveguides and the other end of the fiber is connected to the Optical Spectrum Analyzer (OSA, YOKOGAWA AQ6370C). The optical coupling is achieved by careful alignment of the fibers and the AWG chip. The characterization setup is illustrated in Fig.~\ref{fig:Characterization_stage}. The output fiber is coupled to all the output channels of AWG one by one and the transmission response of each channel is recorded. Similarly, the transmission response of the straight reference waveguide is obtained by coupling the fibers to it. The reference waveguide transmission response is used to normalize the AWG response to obtain the true on-chip response. The normalization process makes the characterization independent of the input intensity and fiber to chip coupling losses. The broadband light source is measured to be steady within 0.05 dB (within 1\%) as a function of time.

\section{Results}
We fabricated the AWG devices as described in the previous section and characterized them. For the first AWG device we used 2.8 $\mu$m width for waveguides and 0.1 $\mu$m height. There are three transmission responses that are important: 

\noindent
\textbf{(1) Transmission from PM fiber to PM fiber}: This defines the input power that is fed to the AWG chip. 
\\*\textbf{(2) Transmission response of the reference waveguide}: This incorporates the fiber to waveguide coupling efficiency (at input and output side) and the propagation loss of the waveguide
\\*\textbf{(3) On-chip AWG throughput}: This is the efficiency of the AWG structure and thus incorporates loss due to curvature of the waveguides, insertion loss between the waveguides and free propagation region (FPR) interfaces and additional propagation loss due to excess length of waveguides in the array. 

\noindent
In this section, we will present the results from our first AWG and further improvement in the overall throughput by adding mode transforming taper between waveguide and fiber and tweaking the waveguide dimensions in the second AWG.  

\subsection{First AWG: Results with Transverse Electric (TE) polarization}
\begin{figure} [ht]
   \begin{center}
   \begin{tabular}{c} 
   \includegraphics[height=5.5cm]{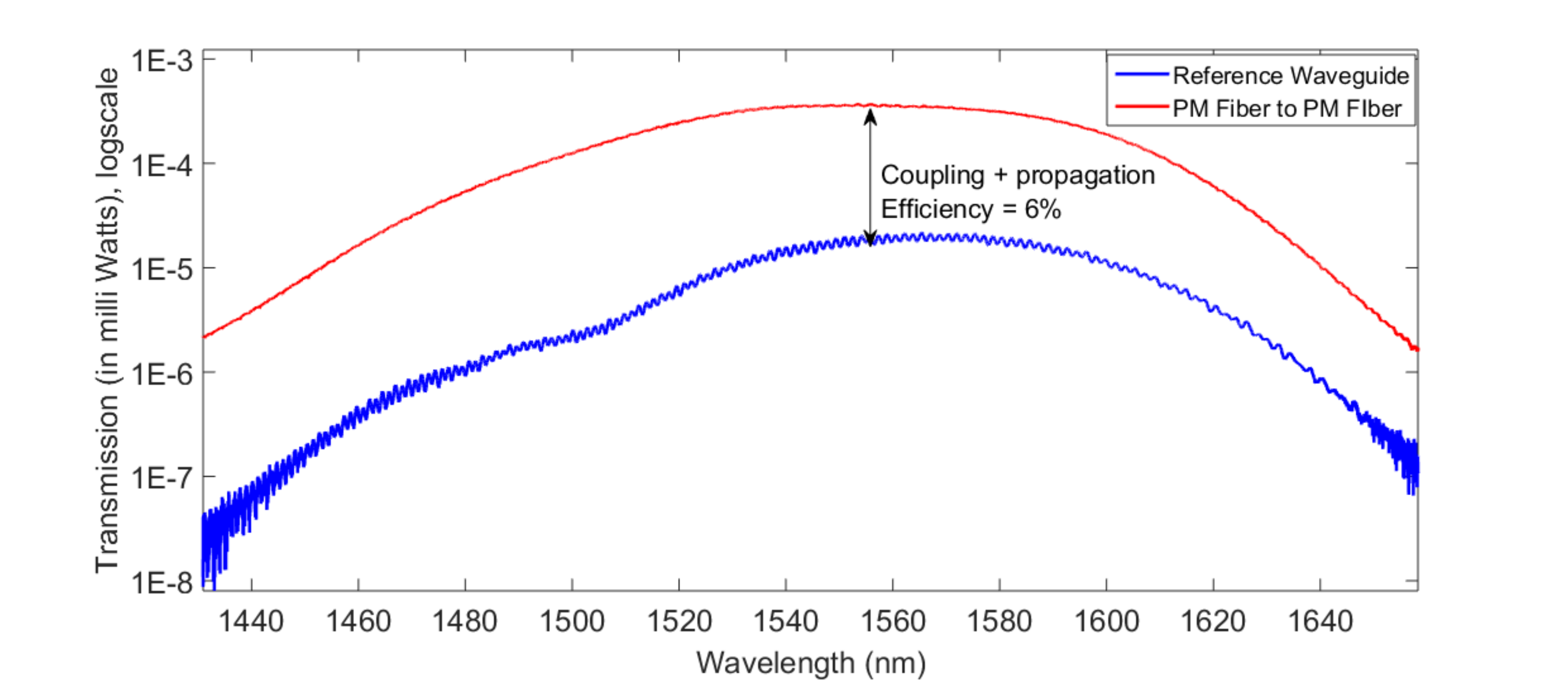}
   \end{tabular}
   \end{center}
   \caption[Reference_Astro] 
   { \label{fig:Reference_Astro} 
 Comparison of PM fiber to PM fiber transmission response (red) with reference waveguide transmission response (blue) for TE polarization for 1st AWG. Most of the loss is due to fiber to waveguide coupling loss.}
   \end{figure}

Fig.~\ref{fig:Reference_Astro} shows PM fiber to PM fiber transmission response and reference waveguide transmission (for TE polarization). Fiber to fiber response is equivalent to input light intensity because the fiber attenuation is very small (\textless 1.0dB/km) [15]. The propagation loss through the straight waveguide is about 25\% as measured separately using waveguides of different lengths. Therefore, most of the loss in Fig.~\ref{fig:Reference_Astro} is contributed by fiber to waveguide coupling loss.

\begin{figure} [ht]
   \begin{center}
   \begin{tabular}{c} 
   \includegraphics[height=5.5cm]{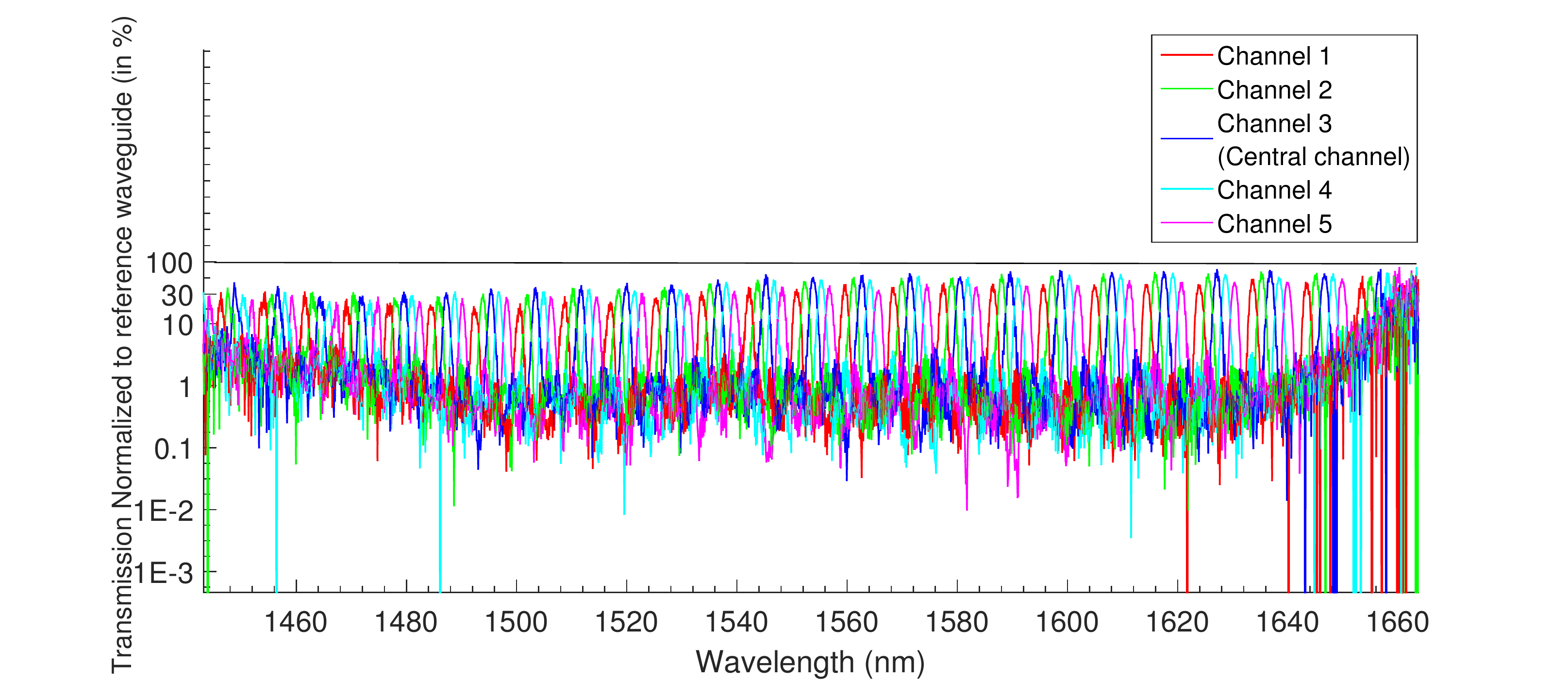}
   \end{tabular}
   \end{center}
   \caption[Full_response] 
   { \label{fig:Full_response} 
   TE polarization normalized transmission response of all 5 output channels of AWG in the wavelength range of 1450 to 1660 nm}
   \end{figure}

\begin{figure} [ht]
   \begin{center}
   \begin{tabular}{c} 
   \includegraphics[height=5.5cm]{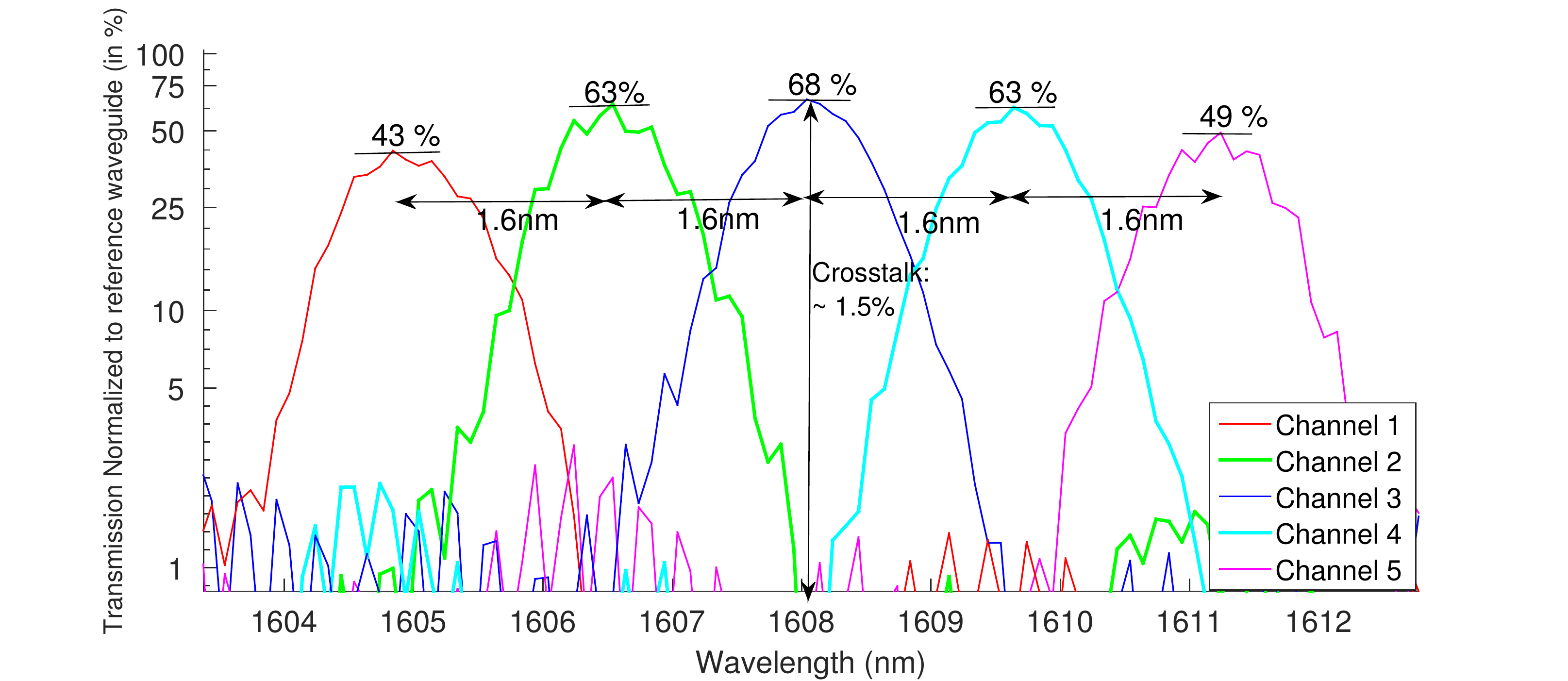}
   \end{tabular}
   \end{center}
   \caption[Partial_response] 
   { \label{fig:Partial_response} 
TE mode transmission response of AWG in range 1604 nm to 1612 nm. Center to center wavelength separation between adjacent channels as well as peak values of normalized transmission for all channels are shown. The small ripples on the peaks are due to inherent spatial variation in the refractive index of the material leading to small phase variations, ultimately causing ripples in the transmission. }
   \end{figure}

The normalized transmission response of all 5 AWG channels for TE polarization is shown in Fig.~\ref{fig:Full_response} (complete) and Fig.~\ref{fig:Partial_response} (around 1600 nm). The AWG transmission is normalized to the reference waveguide transmission to separate the fiber to waveguide coupling losses from on-chip losses. There are 23 spectral orders in the span of 1450 –-- 1650 nm. It is clear from this figure how different wavelengths are focused at different channels (i.e. spatial locations). It can be seen that the central channels in all spectral orders have least loss and the loss increases as we move away from central channel. It is noteworthy that the normalized response is still fairly uniform (within ~20\%) over a large wavelength range of 1450 -- 1650 nm. Also, the contrast ratio is about 2\% for 1500-1640 nm range. This confirms the utility of the preliminary AWG in 1450 -- 1650 nm band for TE polarization. Also, the results indicate that the fabrication recipe and critical parameters used for fabrication are reasonable. This sets the background for the next versions of AWG. Table~\ref{tab:Spectral_parameters} summarizes the variation of key spectral parameters as a function of wavelength. 

\begin{table}[ht]
\caption{Key Spectral parameters of the TE polarization transmission response for the first AWG} 
\label{tab:Spectral_parameters}
\begin{center}       
\begin{tabular}{|c|c|c|c|} 
\hline
\rule[-1ex]{0pt}{3.5ex}   \boldmath$\lambda$ \textbf{(nm)} & \begin{tabular}[t]{@{}c@{}}\textbf{Full width at half}\\\textbf{maximum (FWHM) in nm}\end{tabular} & \begin{tabular}[t]{@{}c@{}}\textbf{Free Spectral Range}\\\textbf{(nm)}\end{tabular} & \begin{tabular}[t]{@{}c@{}}\textbf{Resolving Power}\\\textbf{(R = \boldmath$\lambda$/FWHM)}\end{tabular} \\
\hline
\rule[-1ex]{0pt}{3.5ex}  1450 & 1 & 7.8 & 1450   \\
\hline
\rule[-1ex]{0pt}{3.5ex}  1500 & 0.8 &8.2 & 1875 \\
\hline
\rule[-1ex]{0pt}{3.5ex}  1550 & 1 & 8.5 & 1550  \\
\hline
\rule[-1ex]{0pt}{3.5ex}  1600 & 1 & 9.5 & 1600  \\
\hline 
\rule[-1ex]{0pt}{3.5ex}  1650 & 1.2 & 10 & 1375  \\
\hline

\end{tabular}
\end{center}
\end{table}
\noindent

\subsection{Second AWG: Improvements with tapered geometry}
\begin{figure} [ht]
   \begin{center}
   \begin{tabular}{c} 
   \includegraphics[height=6.7cm]{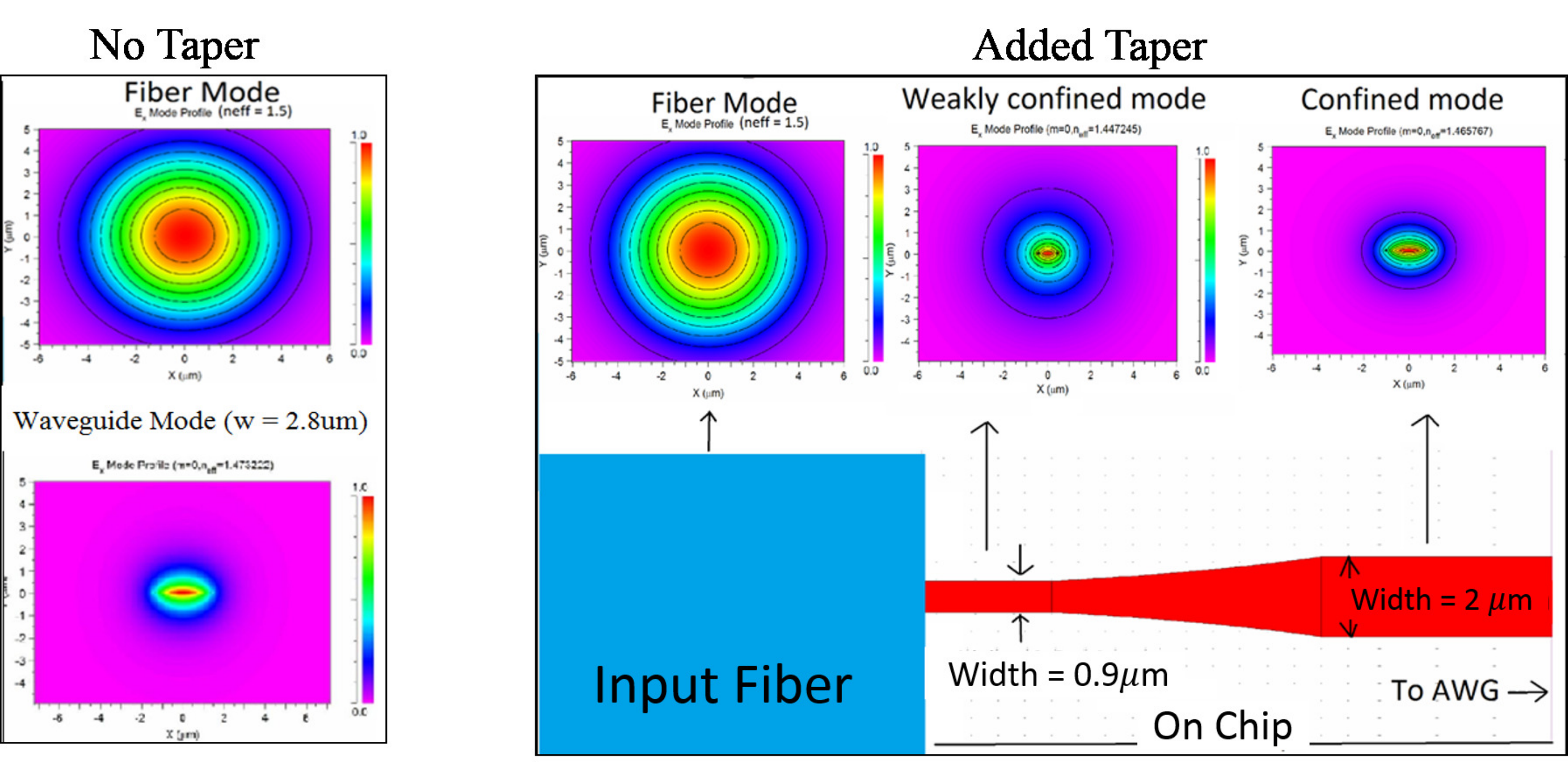}
   \end{tabular}
   \end{center}
   \caption[Taper_mode] 
   { \label{fig:Taper_mode} 
Illustration of the concept of tapered modification of mode index. The taper geometry (top view) is exponential. From left to right, the mode size slowly decreases before finally converging to the mode profile of the waveguide.}
   \end{figure}

As explained earlier, the most important contribution to the overall loss comes from fiber to waveguide coupling loss. The coupling loss occurs due to mismatch between mode profiles of the fiber and the waveguide. One way to solve this problem is using narrow waveguides which tend have very large mode size (squeezed-out modes). But due to their weakly confined modes, we cannot directly use them in AWG. Therefore, we use narrow waveguides at the coupling ends and then lateral adiabatic tapers to slowly increase the waveguide width to the desired value. Then this wide waveguide is used for the AWG structure. This concept is illustrated in Fig.~\ref{fig:Taper_mode}. 

\begin{figure} [ht]
   \begin{center}
   \begin{tabular}{c} 
   \includegraphics[height=16.6cm]{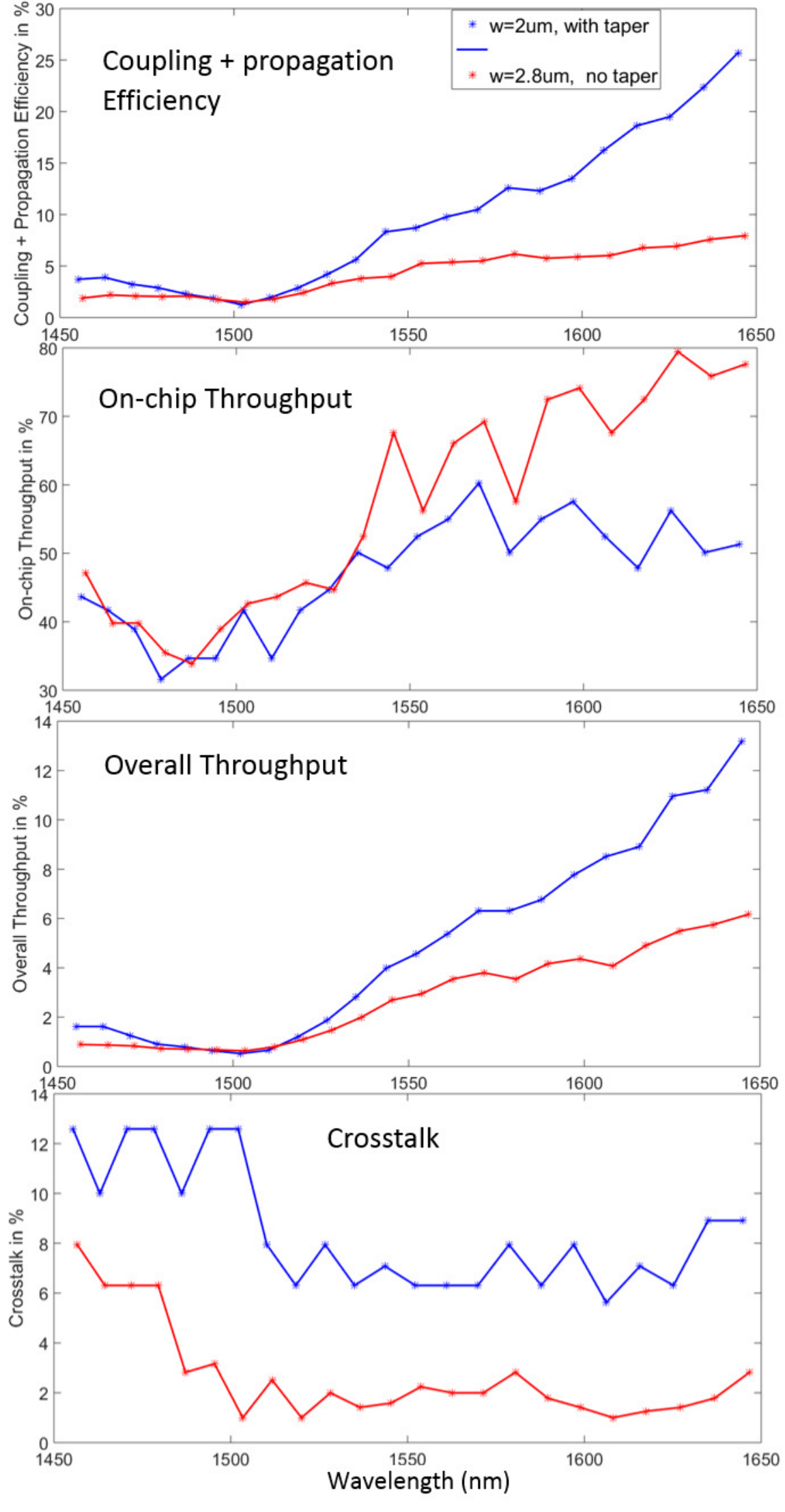}
   \end{tabular}
   \end{center}
   \caption[AWG_comparison] 
   { \label{fig:AWG_comparison} 
 Summary of comparison between measured performance of first (w = 2.8 $\mu$m AWG, shown in red) and second (w = 2 $\mu$m AWG, shown in blue) AWG design. Clearly, the overall throughput (box 3) has significantly increased in the second design, mainly due to better coupling (box 1), but the on-chip performance (box 2) has slightly declined. The lower overall throughput in 1450 -- 1550 nm range is due to unwanted Si-H bond in the SiO$_2$ cladding. This problem is addressed in Section~\ref{Future_work}. Crosstalk (box 4) has  slightly increased for 2 $\mu$m AWG, which is mainly due to the weaker confinement of the mode. Here, each point refers to the central wavelength of each order.}
   \end{figure}

To balance the trade-off between using a weakly confined mode for better coupling efficiency and a strongly confined mode for low AWG losses, we designed an AWG with 2 $\mu$m width (as opposed to previous width of 2.8 $\mu$m) and 0.1 $\mu$m thickness (same thickness as previous). In addition to that we added tapered geometry at the fiber coupling ends of the input and output waveguides (as shown in Fig.~\ref{fig:Taper_mode}). Since 2 $\mu$m waveguide has less confined mode than 2.8 $\mu$m, the on-chip throughput slightly decreases (from $\sim$80\% to $\sim$60\%). But at the same time, the coupling efficiency improves significantly, thus improving overall throughput significantly (from $\sim$8\% to $\sim$26\% ). The crosstalk (contrast ratio) has  slightly increased for 2 $\mu$m AWG (from $\sim$2\% to $\sim$6\%), which is mainly due to the weaker confinement of the mode. The measured results for on-chip throughput, overall throughput and crosstalk are summarized in Fig.~\ref{fig:AWG_comparison} for comparison between first and second AWG.

\begin{table}[ht]
\caption{Comparison between target parameters and measured parameters} 
\label{tab:Summary_results}
\begin{center}       
\begin{tabular}{|l|c|c|c|} 
\hline
\rule[-1ex]{0pt}{3.5ex}   \textbf{Parameters} & \begin{tabular}[t]{@{}c@{}}\textbf{Target}\\\textbf{Specifications}\end{tabular} & \begin{tabular}[t]{@{}c@{}}\textbf{1st AWG}\\\textbf{w=2.8\boldmath$\mu$m}\end{tabular} & \begin{tabular}[t]{@{}c@{}}\textbf{2nd AWG}\\\textbf{(w=2\boldmath$\mu$m with taper)}\end{tabular} \\
\hline
\rule[-1ex]{0pt}{3.5ex}  1. On-chip Throughput & 80\%  &$\sim$80\% (Peak)& $\sim$60\% (Peak)\\
\hline
\rule[-1ex]{0pt}{3.5ex} \begin{tabular}[t]{@{}l@{}}2. Peak Overall Throughput\\(including coupling +\\ propagation)\end{tabular} & 15\% & 6\% & 13\%   \\
\hline
\rule[-1ex]{0pt}{3.5ex}  \begin{tabular}[t]{@{}l@{}}3. Resolving Power\\(at 1600 nm)\end{tabular}  & 1500 & 1600 & 1775   \\
\hline
\rule[-1ex]{0pt}{3.5ex}  \begin{tabular}[t]{@{}l@{}}4. Crosstalk\\(Contrast Ratio)\end{tabular} & 1\% & $\sim$2.5\% & $\sim$7\%  \\
\hline
\rule[-1ex]{0pt}{3.5ex}  5. Operating Waveband & H- and (1450-1700nm) &1450 - 1650 nm & 1450 - 1650 nm  \\
\hline 
\rule[-1ex]{0pt}{3.5ex}  6. Free Spectral Range & 9.7 nm (at 1600 nm) & 9.5 nm (at 1600 nm) & 9.2 nm (at 1600 nm)  \\
\hline

\end{tabular}
\end{center}
\end{table}
\noindent

\subsection{Summary}
In summary, we developed preliminary AWG spectrometer devices for H band with performance close to the preliminary target specifications. Most importantly, we achieved a peak on-chip throughput of 80\% in one device and a peak overall throughput of 13\% in the second device using our design and fabrication techniques. The comparison between target specifications and achieved specifications (for TE polarization) is described in Table ~\ref{tab:Summary_results}. The coupling efficiency significantly increased from first AWG to the second due to the use of taper geometry. The on-chip performance slightly declined from first to the second due to weaker confinement of the waveguide mode leading to slightly higher bending losses and higher interaction between adjacent waveguides, causing a higher crosstalk. But even with this decline, the overall throughput significantly increased from first AWG to the second, since major source of loss is the coupling loss. Current results are very encouraging but more experiments are required to attain an optimum point between coupling and on-chip throughput. 

\noindent
These results confirm the robustness of our design, fabrication and simulation methods and set the path for further improvements. We have achieved the operating waveband of almost entire H band (1450 -- 1650 nm). To characterize the device beyond 1650 nm we need to use a wider band source in the future. 

\section{Future Work}
\label{Future_work}
The future work will be mainly focused on increasing the throughput at moderate resolution and then going to higher resolution. At present, there are following practical ways to increase the throughput which will be explored in future. 

\noindent
\textbf{Using better taper geometries}: Currently, we are using 0.9 $\mu$m to 2 $\mu$m taper. Further reducing the initial width will make the mode size larger, thus matching the fiber mode profile better. Our beam propagation simulations using BeamPROP software\cite{BeamPROP} (using Beam Propagation Method), indicate that a starting width of 0.6 $\mu$m comes closest to the fiber mode profile. Therefore, it is expected to give a higher coupling efficiency. 

\noindent
\textbf{Increasing throughput in 1450 –- 1550 nm range using annealing}: The increased attenuation in the range 1450 to 1550 nm (see Fig.~\ref{fig:AWG_comparison}) is due to presence of unwanted Si-H bond in the cladding SiO$_2$\cite{henry1987low} that is deposited using PECVD (Plasma Enhanced Chemical Vapor Deposition) in the last step of fabrication (see Fig.~\ref{fig:Fabrication_process}). This bond absorbs in NIR with peak absorption at 1400 and 1520 nm. Annealing the sample in a certain heating profile to 1200$^0$C after PECVD step is a potential solution to alleviate this problem\cite{henry1987low} by liberating the hydrogen.     

\noindent
\textbf{Improving on-chip throughput}: Reducing the thickness of the waveguides has been demonstrated as a way to achieve better on-chip throughput\cite{dai2011low}. With a reduced thickness, the width of the waveguides need to be increased to maintain the same confinement of the mode. With a wide waveguide, the mismatch between the waveguide and the free propagation region on the chip is reduced, thus improving the on-chip throughput. Also, due to reduced height, the sidewall scattering area is reduced, thus reducing the scattering loss.

\noindent
Apart from these improvements, it is important to improve the Transverse Magnetic (TM) polarization response of the AWG since most of the astronomical light to be observed with this device is largely going to be unpolarized. Making a polarization-independent AWG over such a wide band is a challenging problem due to the inherent birefringence of the waveguide geometry. One way to solve this problem is by using a polarization splitter to feed TE light to one AWG and TM light to another AWG (specifically designed for TM mode). The other way will be to develop a geometry of the waveguides that is polarization independent over a wide band. A square-shaped geometry is more suited for this purpose, but there are other constraints such as single-mode (fundamental TE and TM) propagation condition, sidewall roughness for deeper etches and wideband performance. Therefore, currently it is challenging to achieve.      

\acknowledgments 

The authors thank the University of Maryland NanoCenter and Fablab for providing all the fabrication equipment. In particular, we thank Jonathan Hummel, Tom Loughran and Mark Lecates for fabrication process training and advice. The authors thank Prof. Stuart Vogel for his suggestions. The authors acknowledge the financial support for this project from the W. M. Keck Foundation.

\bibliography{report} 
\bibliographystyle{spiebib} 

\end{document}